\documentclass[aps,pra,amsmath,amssymb,twocolumn,superscriptaddress,showpacs]{revtex4-1}
\usepackage{ulem}
\usepackage{amsmath,amssymb}
\usepackage{graphicx}	
\mathchardef\mhyphen="2D

\def\be{\begin{eqnarray}}   
\def\ee{\end{eqnarray}}
\def\vecb{\boldsymbol}

\begin{document}

\author{Wenwen~Mao}
\affiliation 
{Max Planck Institute for the Structure and Dynamics of Matter, Luruper Chaussee 149, 22761 Hamburg, Germany}

\author{Angel~Rubio}
\affiliation 
{Max Planck Institute for the Structure and Dynamics of Matter, Luruper Chaussee 149, 22761 Hamburg, Germany}
\affiliation 
{Center for Computational Quantum Physics (CCQ), Flatiron Institute, 162 Fifth Avenue, New York, NY
10010, USA}

\author{Shunsuke~A.~Sato}
\email{ssato@ccs.tsukuba.ac.jp}
\affiliation 
{Center for Computational Sciences, University of Tsukuba, Tsukuba 305-8577, Japan}
\affiliation 
{Max Planck Institute for the Structure and Dynamics of Matter, Luruper Chaussee 149, 22761 Hamburg, Germany}

\title{Terahertz-induced high-order harmonic generation and nonlinear charge transport in graphene}

\begin{abstract}
We theoretically study the THz-induced high-order harmonic generation (HHG) and nonlinear electric transport in graphene based on the quantum master equation with the relaxation time approximation. To obtain microscopic insight into the phenomena, we compare the results of the fully dynamical calculations with those under a quasi-static approximation, where the electronic system is approximated as a nonequilibrium steady state. As a result, we find that the THz-induced electron dynamics in graphene can be accurately modeled with the nonequilibrium steady-state at each instance. The population distribution analysis further clarifies that the THz-induced HHG in graphene originates from the reduction of effective conductivity due to a large displacement of electrons in the Brillouin zone. By comparing the present nonequilibrium picture with a thermodynamic picture, we explore the role of the nonequilibrium nature of electron dynamics on the extremely nonlinear optical and transport phenomena in graphene.
\end{abstract}

\maketitle
\section{Introduction \label{sec:intro}}

High-order harmonic generation (HHG) is an extreme photon-upconversion process via strongly nonlinear light-matter interactions, and it has been intensively studied in gas systems \cite{PhysRevLett.68.3535,PhysRevLett.70.1599,PhysRevA.49.2117}, enabling the generation of attosecond laser pulses and opening a novel avenue to study ultrafast electron dynamics in the time domain \cite{RevModPhys.81.163,Goulielmakis2010,doi:10.1126/science.1260311,doi:10.1126/science.aag1268}. After the discovery of HHG in ZnO crystal \cite{Ghimire2011}, the HHG in extended systems has been attracting much interest as it may further contribute to the development of novel light sources \cite{Ghimire2019}. Among various materials, HHG in graphene has been intensively studied both theoretically \cite{wright2009strong,ishikawa2010nonlinear,al2014high,al2015optimizing,sorngaard2013high,chizhova2017high} and experimentally \cite{yoshikawa2017high,cox2017plasmon} as graphene has a unique electronic structure, Dirac cones. Recently, the HHG in graphene has been investigated in the THz regime \cite{hafez2018extremely,kovalev2021electrical}. Furthermore, the field-induced transparency of graphene has been investigated as yet another intriguing nonlinear optical effect in the THz regime \cite{Hwang2013,Paul_2013,doi:10.1063/1.4902999}. These nonlinear optical effects have been addressed based on the reduction of the electric conductivity with the thermodynamic model \cite{mics2015thermodynamic,kovalev2021electrical}. However, the microscopic mechanism of these nonlinear effects still has not been understood based on the nonequilibrium quantum dynamics under dissipation beyond the phenomenological treatment.

To develop the microscopic understanding of the physical mechanism of the HHG in graphene in the THz regime, we investigate the nonequilibrium electron dynamics in graphene with the quantum master equation. We simulate the THz-induced electron dynamics under the dissipation and compare it with a nonequilibrium steady state under a static field. As a result, we find that the THz-induced electron dynamics in graphene can be well described, at each instance, with the nonequilibrium steady state. Furthermore, the nonequilibrium simulation clarifies that the effective electric conductivity of graphene is reduced due to the depletion of effective carriers, resulting in the nonlinear current and the HHG. In this work, we further compare the present nonequilibrium description of electron dynamics in graphene and the recently developed thermodynamic model \cite{mics2015thermodynamic} in order to clarify the role of the nonequilibrium nature of dynamics in the nonlinear optical phenomena in graphene.

The paper is organized as follows. In Sec.~\ref{sec:method}, we first describe theoretical methods to study the light-induced electron dynamics in graphene based on the quantum master equation. In Sec.~\ref{sec:result}, we investigate the THz-induced HHG in graphene with the method described in Sec.~\ref{sec:method}. We further analyze the microscopic mechanism of HHG with the quasi-static approximation and the population distribution in the Brillouin zone. In Sec.~\ref{sec:thermo}, we elucidate the role of the nonequilibrium nature of THz-induced electron dynamics by comparing the nonequilibrium picture in the present work and the thermodynamic picture in the previous work \cite{mics2015thermodynamic}. Finally, our findings are summarized in Sec.~\ref{sec:summary}.
\section{Methods \label{sec:method}}
\subsection{Theoretical modeling \label{subsec:model}}

In this work, we describe the light-induced electron dynamics in graphene with the following quantum master equation \cite{sato2019light,sato2019microscopic,sato2021high,sato2021nonlinear}:
\begin{equation}
\frac{\mathrm{d}}{\mathrm{d}t}\rho_{\boldsymbol{k}}(t) = \frac{1}{i \hbar}	\left[ H_{\boldsymbol{k}+e\boldsymbol{A}(t)/\hbar}, \rho_{\boldsymbol{k}} (t)) \right] + 	
\hat{D}\left[ \rho_{\boldsymbol{k}} (t)) \right],
\label{eqn:masterequation}
\end{equation}
where $\boldsymbol{k}$ is the Bloch wavevector, $\rho_{\vecb k}(t)$ is the reduced density matrix at $\vecb k$, $\vecb A (t)$ is a spatially-uniform vector potential related to applied electric fields as $\vecb{A} = -\int^t_{-\infty}dt' \vecb{E}(t')$. The time-dependent Hamiltonian, $H_{\vecb{k}+e\vecb{A}(t)/\hbar}$, is constructed with the Peierls substitution~\cite{hofstadter1976energy}, simply replacing the Bloch wavevector, $\vecb k$, with the shifted wavevector, $\vecb k + e\vecb A(t)/\hbar$, in the static Hamiltonian $H_{\vecb k}$. To describe the electronic structure of graphene, we employ the following tight-binding Hamiltonian \cite{neto2009electronic}:
\begin{equation}
H_{\boldsymbol{k}}=\left(\begin{array}{cc}
0 & t_{0} f(\boldsymbol{k}) \\
t_{0} f(\boldsymbol{k})^{*} & 0
\end{array}\right),
\label{eqn:TBhamiltonian}
\end{equation}
where $t_0$ is the nearest-neighbor hopping, and $f(\boldsymbol{k})$ is given by $f(\boldsymbol{k})=e^{i \boldsymbol{k} \cdot \vecb{\delta}_{1}}+e^{i \boldsymbol{k} \cdot \vecb{\delta}_{2}}+e^{i \boldsymbol{k} \cdot \vecb{\delta}_{3}}$ with the nearest-neighbor vectors $\vecb \delta_j$~\cite{neto2009electronic}. We set the hopping parameter $t_0$ to $2.8$~eV and the lattice constant $a$ to $1.42$~$\AA$ in accordance with the previous work \cite{neto2009electronic}.

To describe the effect of dissipation, we construct the relaxation operator, $\hat{D}\left[\rho_{\vecb{k}} (t)\right]$, in Eq.~(\ref{eqn:masterequation}) with the relaxation time approximation~\cite{meier1994coherent} with the Houston basis \cite{PhysRev.57.184,PhysRevB.33.5494}. The Houston states are eigenstates of the instantaneous Hamiltonian: $H_{\vecb{k}+e\vecb{A}(t)/\hbar} |u^H_{b\vecb k}(t)\rangle = \epsilon_{b,\vecb k + e\vecb A(t)/\hbar}|u^H_{b\vecb k}(t)\rangle$, where $b$ denotes the band index, valence ($b=v$) or conduction ($b=c$) bands. The reduced density matrix can be expanded with the Houston states as
\begin{align}
  \rho_{\vecb k}(t) = \sum_{bb'}\rho_{bb',\vecb k}(t)|u^H_{b\vecb k}(t)\rangle \langle u^H_{b'\vecb k}(t)|,
\end{align}
where $\rho_{bb',\vecb k}(t)$ are the expansion coefficients. On the basis of the Houston state expansion, we define the relaxation operator \cite{sato2019microscopic} as
\begin{widetext}
\begin{align}
  \hat D\left [\rho_{\vecb k}(t) \right ]=-\sum_{b}\frac{\rho_{bb,\vecb k}(t)-f^{FD}\left (\epsilon_{b,\vecb k+e\vecb A(t)/\hbar},T_e,\mu \right)}{T_1}|u^H_{b\vecb k}(t)\rangle \langle u^H_{b\vecb k}(t)| 
  -\sum_{b\neq b'} \frac{\rho_{bb',\vecb k}(t)}{T_2}|u^H_{b\vecb k}(t)\rangle \langle u^H_{b'\vecb k}(t)|,
\label{eqn:relaxation}
\end{align}
\end{widetext}
where $T_1$ is the longitudinal relaxation time, $T_2$ is the transverse relaxation time, and $f^{\mathrm{FD}}(\epsilon)$ is the Fermi--Dirac distribution
\begin{align}
f^{\mathrm{FD}}(\epsilon, T_e, \mu)=\frac{1}{e^{(\epsilon-\mu)/k_BT_e}+1}.
\label{eq:fd-dist}
\end{align}
Here, $\mu$ is the chemical potential, and $T_e$ is the electron temperature.
In this work, we set the longitudinal relaxation time $T_1$ to $100$~fs and the transverse relaxation time $T_2$ to $20$~fs in accordance with the previous works \cite{sato2021nonlinear,sato2021high,sato2019light,sato2019microscopic}. The electron temperature $T_e$ is set to $300$~K unless stated otherwise. The chemical potential $\mu$ is treated as a tunable parameter to study the effect of doping.

By employing the time-dependent density matrix $\rho_{\vecb k}(t)$ evolved with Eq.~(\ref{eqn:masterequation}), we compute the electric current as
\begin{eqnarray}
  \vecb{J}(t)=\frac{2}{(2\pi)^2} \int d\vecb k \mathrm{Tr}\left[\hat{\boldsymbol{J}}_{\boldsymbol{k}}(t)\rho_{\boldsymbol{k}}(t)\right],
\label{eqn:totalcurrent}
\end{eqnarray}
where $\hat{\boldsymbol{J}}_{\boldsymbol{k}}(t)$ is the current operator defined as
\begin{align}
\hat{\boldsymbol J}_{\boldsymbol{k}}(t) = -\frac{\partial H(\boldsymbol{k}+e\boldsymbol{A}(t)/\hbar)}{\partial \boldsymbol A(t)}.
\end{align}
By analyzing the current induced by electric fields, we further investigate the high-order harmonic generation and nonlinear transport properties of graphene.

\section{Results \label{sec:result}}

In this section, we study the microscopic mechanism behind the THz-induced high-order harmonic generation in graphene. We first investigate the THz-induced electron dynamics in graphene with fully dynamical simulations based on the quantum master equation, Eq.~(\ref{eqn:masterequation}). Then, we introduce a quasi-static approximation to analyze the THz-induced electron dynamics, revisiting the nonlinear electric transport and field-induced transparency of graphene. Furthermore, we compare a nonequilibrium steady state realized in the quasi-static picture with the recently developed thermodynamic model \cite{mics2015thermodynamic} in order to clarify the nonequilibrium mechanism behind nonlinear optical and transport phenomena in graphene in the THz regime.

\subsection{Fully dynamical simulations for high-order harmonic generation in graphene \label{subsec:dynamical_simulation}}

We first perform the electron dynamics simulation with Eq.~(\ref{eqn:masterequation}) by using a linearly polarized laser pulse in order to analyze the high-order harmonic generation in graphene. For this purpose, we employ the following form for the applied vector potential
\begin{align}
  \vecb A(t) = -\frac{E_0}{\omega_0}\vecb{e_x} \sin(\omega_0 t)\cos^4 \left (\frac{\pi}{T_\mathrm{full}} t \right)
  \label{eqn:laser_pulse}
\end{align}
in the domain $-T_\mathrm{full}/2<t<T_\mathrm{full}/2$ and zero outside. In accordance with the previous experiment~\cite{hafez2018extremely}, we set the peak field strength $E_0$ to $8.5$~MV/m, the mean photon-energy $\hbar \omega_0$ to $1.2407$~meV, and the pulse duration $T_{\mathrm{full}}$ to $40$~ps. The direction of the electric field $\vecb e_x$ is set to $\Gamma$--$M$ direction.

We compute the induced electric current, $\vecb J(t)$, under the field given by Eq.~(\ref{eqn:laser_pulse}). Then, we apply the Fourier transform to the current in order to evaluate the high-order harmonics spectrum as
\begin{align}
I_{\mathrm{HHG}}(\omega)\sim \omega^2 \left | \int^{\infty}_{-\infty} dt J(t) e^{i\omega t} \right |^2.
\label{eqn:spectrum}
\end{align}

\begin{figure}[htb]
\includegraphics[width=0.9\linewidth]{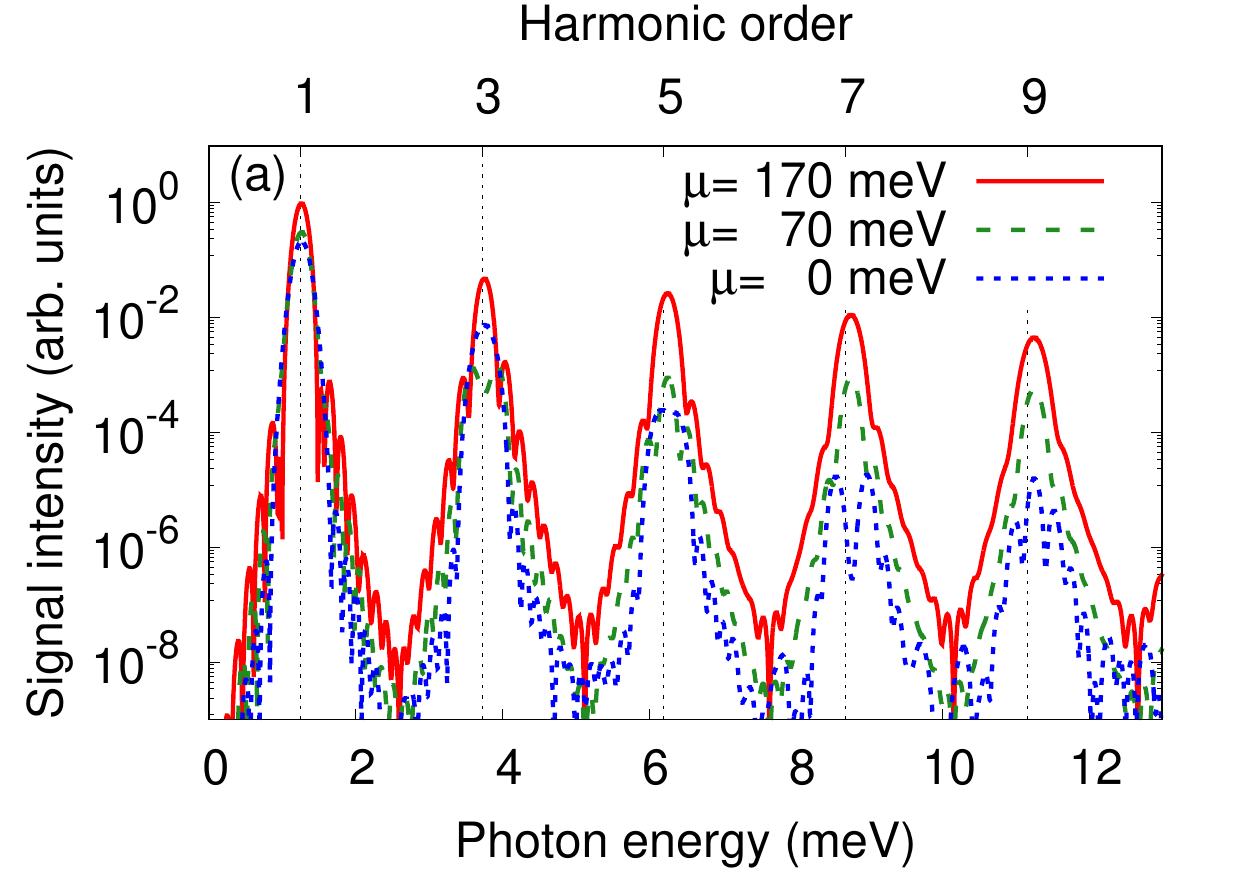}
\includegraphics[width=0.9\linewidth]{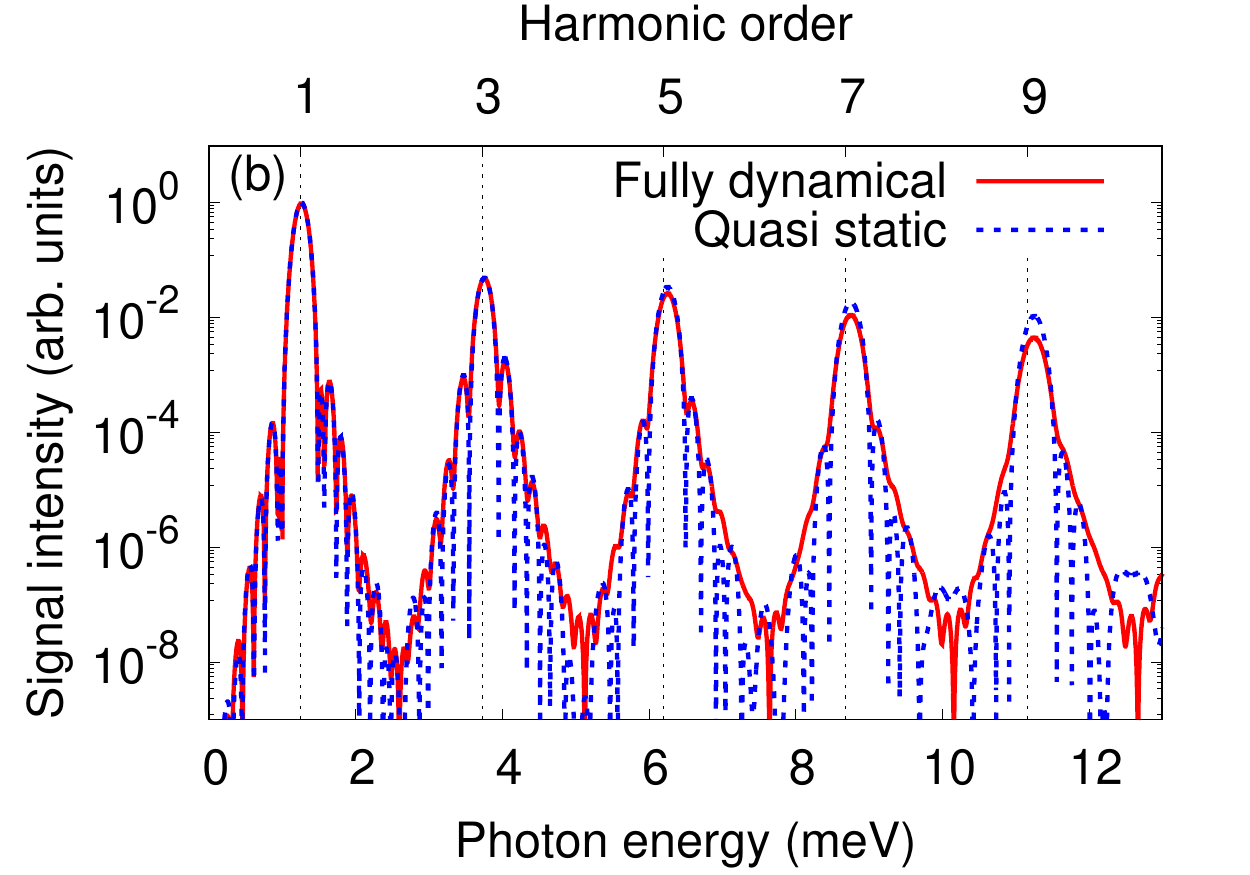}
\caption{\label{fig:hhg_mu} 
(a) Computed harmonic spectra $I_{\mathrm{HHG}}(\omega)$ with Eq.~(\ref{eqn:spectrum}) for different chemical potentials, $\mu = 0$, $70$ and $170$~meV. (b) Comparison of the HHG spectra computed with the fully dynamical simulations in Sec.~\ref{subsec:dynamical_simulation} and the quasi-static approximation Sec.~\ref{subsec:quasi-static}. Here, the chemical potential is set to $\mu =170$~meV.}
\end{figure}

Figure~\ref{fig:hhg_mu}~(a) shows the computed high-order harmonic spectra, $I_{\mathrm{HHG}}(\omega)$, for different chemical potentials $\mu$. For each chemical potential, clear harmonic peaks are observed. The intensities of emitted harmonics increase with the increase in the chemical potential. These results are consistent with the observation in the recent experiment \cite{kovalev2021electrical}, where the emitted harmonics intensity increases with the increase in the gate voltage. In the previous work, the THz-induced high-order harmonic generation in graphene was interpreted on the basis of the thermodynamic picture \cite{mics2015thermodynamic}. In this work, we aim to develop a comprehensive microscopic understanding of the THz-induced nonlinear phenomena by taking into account the nonequilibrium nature of electron dynamics in the description of light-matter interactions.

\subsection{Quasi-static approximation for THz-induced electron dynamics in graphene \label{subsec:quasi-static}}

To develop a microscopic understanding of the THz-induced high-order harmonic generation in graphene, we introduce a quasi-static picture to describe the induced electron dynamics \cite{sato2021nonlinear}. Here, we assume that the THz-field varies so slowly that the electronic system can be well described with a nonequilibrium steady state at each time under the balance between the field-induced excitation and the relaxation. This assumption becomes accurate when the mean frequency of the THz field is much smaller than the intrinsic relaxation rates, $1/T_1$ and $1/T_2$.

For practical analysis with the quasi-static approximation, we first evaluate the electric current of a nonequilibrium steady state under a static electric field, $\vecb E(t)=E_0 \vecb e_x $ as
\begin{align}
  \vecb J_S(E_0) = \lim_{t\rightarrow \infty}\frac{2}{(2\pi)^2}\int d\vecb k \mathrm{Tr}\left[\hat{\boldsymbol{J}}_{\boldsymbol{k}}(t)\rho_{\boldsymbol{k}}(t)\right].
  \label{eq:steady-current}
\end{align}
Here, the electron dynamics are computed under a static field, $\vecb{A}(t)=-E_0 \vecb{e}_x t$. The electronic system reaches a nonequilibrium steady state after sufficient time due to the balance between the field-induced excitation and the relaxation (see Appendix~\ref{appendix:qs-picture} for details). With the relation between the current and the field in Eq.~(\ref{eq:steady-current}), we approximate the field-induced current $\vecb J(t)$ by the steady-state current with the instantaneous electric field as $\vecb J(t)\approx \vecb J_S\left( \vecb E(t) \right)$.

To assess the accuracy of the quasi-static approximation, we computed the high-order harmonic generation spectrum $I_{\mathrm{HHG}}(\omega)$ with the approximated current, $\vecb J_S\left( \vecb E(t) \right)$. Figure~\ref{fig:hhg_mu}~(b) shows the computed spectrum $I_{\mathrm{HHG}}(\omega)$ with the quasi-static approximation by setting $\mu$ to $170$~meV. For comparison, the corresponding result of the fully dynamical calculation is also shown. As seen from the figure, the result of the quasi-static approximation accurately reproduces that of the fully dynamical calculation. Hence, we confirm that the quasi-static approximation can well describe the electron dynamics in graphene under THz fields. This indicates that the microscopic mechanism of the THz-induced HHG in graphene can be developed on the basis of the nonequilibrium steady state under the balance between the field-induced excitation and the relaxation. Note that the quasi-static approximation becomes less accurate for the higher-order harmonics due to the fast component of the dynamics that cannot be well captured by the quasi-static picture.

\subsection{Nonlinear electric conductivity of graphene \label{subsec:nonlinear-sigma}}

Having established the quasi-static picture of THz-induced electron dynamics in graphene, we then study the nonlinear electric conductivity in a static regime in order to develop microscopic insight into the THz-induced HHG. For this purpose, we first define the intraband component of the steady-state current in Eq.~(\ref{eq:steady-current}) as
\begin{align}
\boldsymbol{J}^{\mathrm{intra}}_S(E_0)&=\sum_{b=v,c} \lim\limits_{t\rightarrow \infty}\frac{(-2)}{(2\pi)^2}
\frac{e}{\hbar} \nonumber \\
&\times 
\int d\boldsymbol{k}\frac{\partial\epsilon_{b,\boldsymbol{k}+e\boldsymbol{A}(t)/\hbar}}{\partial\boldsymbol{k}} n_{b,\boldsymbol{k}+e\boldsymbol{A}(t)/\hbar},
\label{eqn:intra-steady-current}
\end{align}
where the band population $n_{b,\boldsymbol{k}+e\boldsymbol{A}(t)/\hbar}$ is defined as $n_{b,\boldsymbol{k}+e\boldsymbol{A}(t)/\hbar}(t)=\langle u^H_{b,\boldsymbol k}(t) |\rho_{\boldsymbol{k}}(t)|u^H_{b,\boldsymbol k}(t) \rangle$ with the instantaneous eigenstates of the Hamiltonian, $|u^H_{b,\boldsymbol k} (t)\rangle$. We then evaluate the effective conductivities from the total steady current $\vecb J_S(E_0)$ and the intraband component $\vecb J^{\mathrm{intra}}_S(E_0)$ as $\sigma(E_0)=\vecb e_x\cdot \vecb J_S(E_0)/E_0$ and $\sigma^{\mathrm{intra}}(E_0)=\vecb e_x\cdot \vecb J^{\mathrm{intra}}_S(E_0)/E_0$, respectively.

\begin{figure}[htb]
\includegraphics[width=0.9 \linewidth]{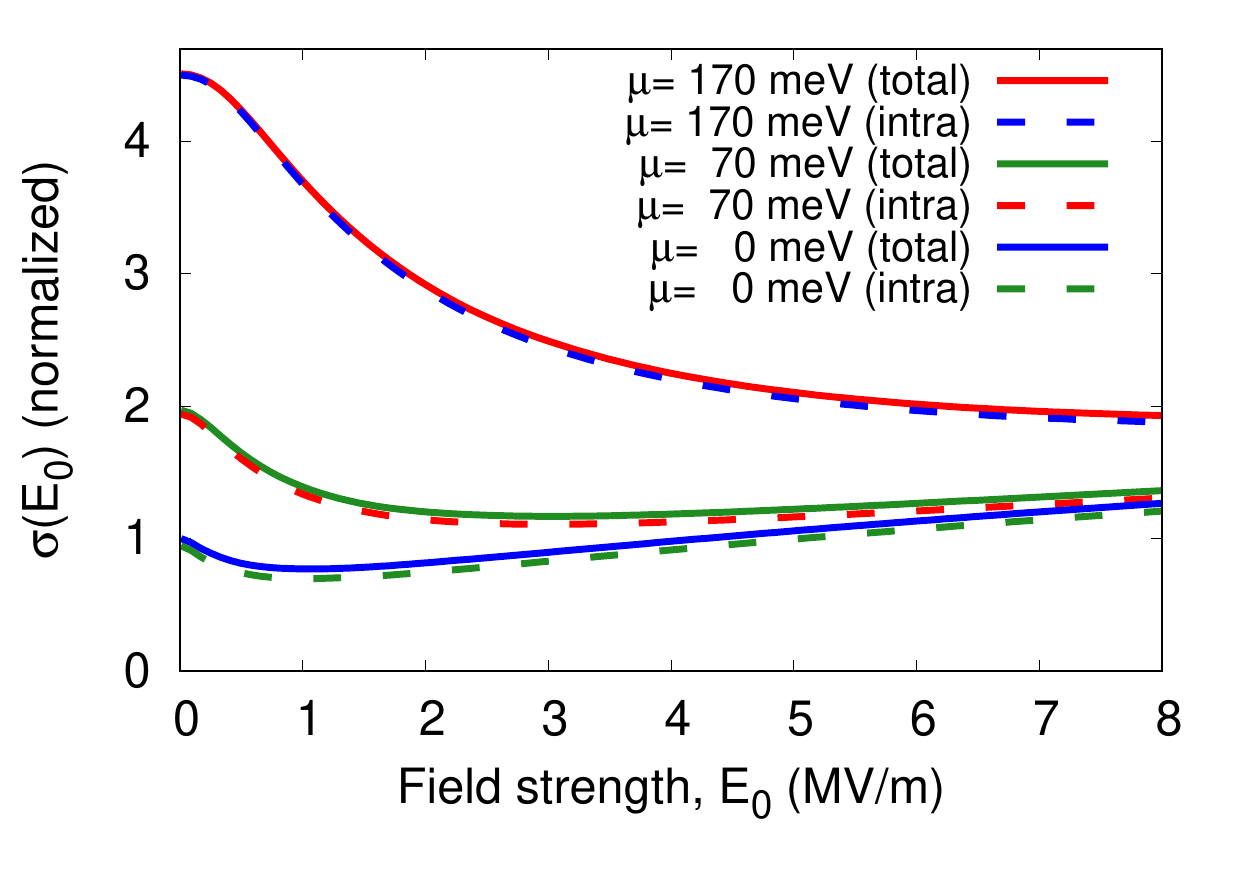}
\caption{\label{fig:conductivity} 
Nonlinear effective conductivities of graphene as a function of the static field strength $E_0$ evaluated with the total currents (solid lines) and intraband currents (dashed lines) for different values of the chemical potential, $\mu = 0$, $70$ and $170$~meV.}
\end{figure}

Figure~\ref{fig:conductivity} shows the computed effective conductivities, $\sigma(E_0)$ and $\sigma^{\mathrm{intra}}(E_0)$, as a function of the applied field strength $E_0$. The results for different chemical potentials, $\mu$, are shown. In Fig.~\ref{fig:conductivity}, the conductivities $\sigma(E_0)$ evaluated with the total steady current $\vecb J_S(E_0)$ are well reproduced with those evaluated with the intraband current $ \vecb J^{\mathrm{intra}}_S(E_0)$ for all investigated field strength $E_0$ and chemical potential $\mu$. Therefore, the charge transport in graphene under static and THz fields is dominated by the intraband current, which is described by the product of the band group velocity and the band population in the Brillouin zone.

As seen from Fig.~\ref{fig:conductivity}, the effective conductivities, $\sigma(E_0)$, are first reduced for all investigated chemical potentials $\mu$ when the field strength increases from zero. The reduction of the conductivity is consistent with the field-induced transparency of graphene \cite{sato2021nonlinear} since the conductivity $\sigma(E_0)$ is directly related to the photoabsorption via Joule heating, $E_{\mathrm{Joule}}=\vecb E_0 \cdot \vecb J_S(E_0)=\sigma(E_0)E^2_0$. Once the field strength becomes even stronger, graphene with relatively small chemical potentials (e.g., $\mu=0$ or $70$~meV) shows the conductivity increase, while graphene with the relatively large chemical potential (e.g., $\mu=170$~meV) keeps showing the conductivity decrease. These results are consistent with the previous theoretical study on the nonlinear transport in graphene with the linear band approximation, $H_{\vecb k}=v_F\left (\sigma_x k_x +\sigma_y k_y \right)$ \cite{sato2021nonlinear}. Since the present work employs a more comprehensive electronic structure in the full Brillouin zone based on the tight-binding model, the low-energy Hamiltonian approximation for the graphene bandstructure in the previous work can be verified on the basis of the present results. In the previous work \cite{sato2021nonlinear}, the decrease of the effective conductivity has been understood by the dispersion of the population imbalance in the Brillouin zone, and the conductivity increase has been understood by the additional carrier injection via the Zener tunneling mechanism. These interpretations can be naturally applied to the present results.

Since the quasi-static approximation well describes the THz-induced electron dynamics, the THz-induced HHG can be interpreted on the basis of the effective conductivities $\sigma(E_0)$ in Fig.~\ref{fig:conductivity}. If the conductivity $\sigma(E_0)$ is independent of the field strength $E_0$, the induced current is always linearly proportional to the field strength, resulting in the absence of harmonics.  Therefore, the emitted harmonics in the quasi-static picture originate from the nonlinearity of the current $\vecb J_S(E_0)$ and the field-strength dependence of the conductivity $\sigma(E_0)$. As seen from Fig.~\ref{fig:conductivity}, the conductivity has a stronger dependence on the field strength for a larger chemical potential, manifesting the significant conductivity reduction with the increase in the field strength. This indicates that the enhancement of the HHG with the chemical-potential shift in Fig.~\ref{fig:hhg_mu} can be understood by the significant reduction of the conductivity with the increase in the field strength at a larger chemical potential. In the previous work \cite{hafez2018extremely,kovalev2021electrical}, the THz-induced HHG in graphene was also interpreted by the reduction of the conductivity but with the thermodynamic model \cite{mics2015thermodynamic}. To understand the role of the nonequilibrium nature in the steady-state, we elucidate a relation of the two models, the nonequilibrium steady-state model, and the thermodynamic model, in the forthcoming section, Sec.~\ref{sec:thermo}.

\begin{figure*}[htb]
\includegraphics[width=0.7\linewidth]{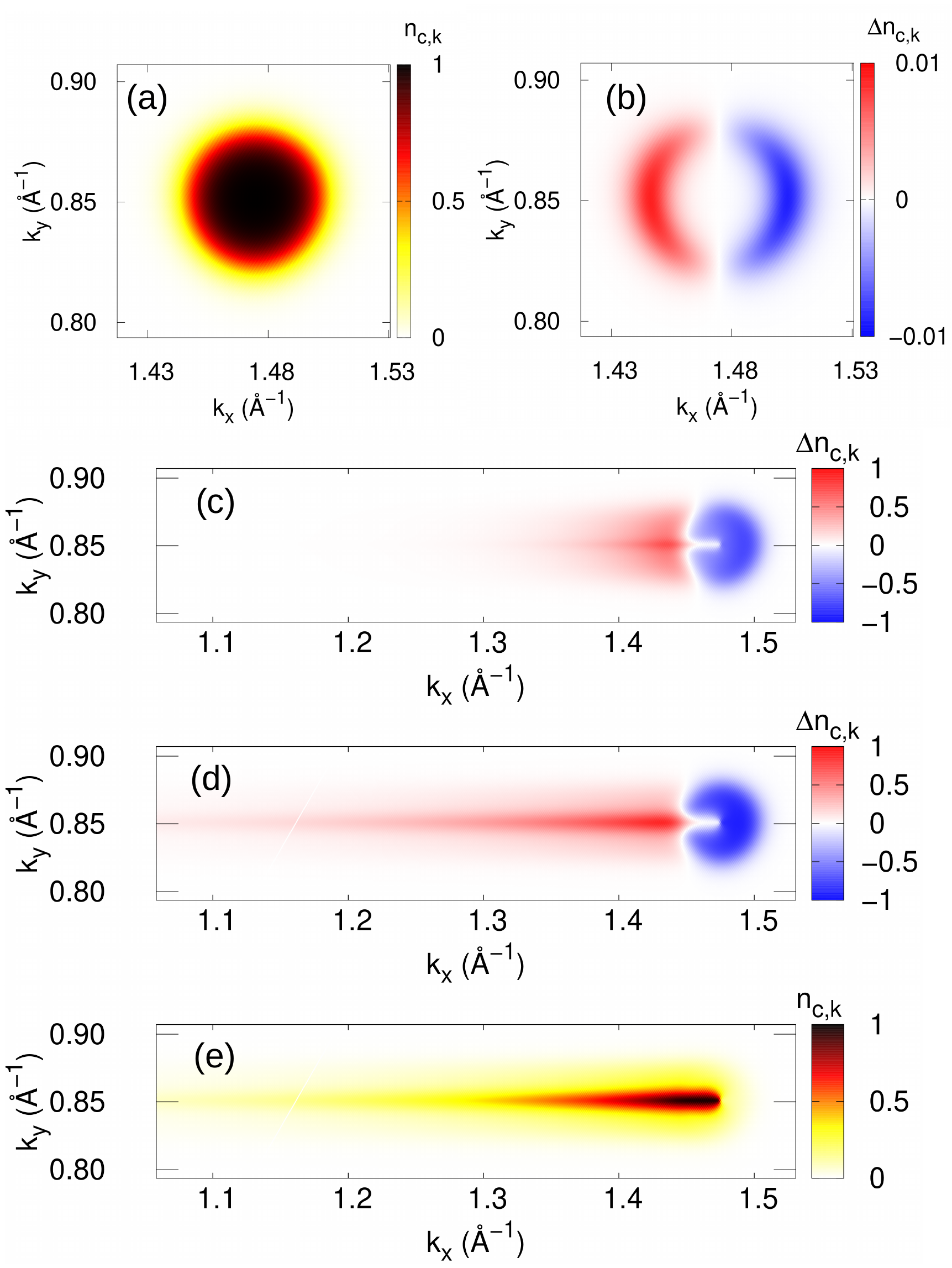}
\caption{\label{fig:pop_f}
(a) The equlibrium population distribution in the conduction band $f^{\mathrm{FD}}(\epsilon_{c,\vecb k})$. (b-d) The field induced conduction population change for different field strengths, (b) $0.01$~MV/m, (c) $3$~MV/m, and (d) $10$~MV/m. (e) The population distribution in the conduction band in the nonequilibrium steady-state under a static field, $E_0=10$~MV/m.}
\end{figure*}

The intraband current in Eq.~(\ref{eqn:intra-steady-current}) consists of the product of the band velocity and population. Since the band velocity is an intrinsic property of material and invariant under the presence of electric fields, the field-induced population modification plays an essential role in the generation of the intraband current. Furthermore, the THz-induced current is dominated by the intraband current as discussed above. To obtain microscopic insight into the THz-induced current, we thus analyze the population distribution in the Brillouin zone under the field. Figure~\ref{fig:pop_f}~(a) shows the equilibrium population distribution in the conduction band, $f^{\mathrm{FD}}(\epsilon_{c,\vecb k})$, around a Dirac point (K point) of graphene: $\vecb k =\frac{2\pi}{\sqrt{3}a}\left(1, \frac{1}{\sqrt{3}}\right)$. Here, the chemical potential $\mu$ is set to $170$~meV. One sees that the equilibrium population is distributed around the Dirac point with circular symmetry as the Dirac cone is partially filled by doped electrons.

We define the field-induced conduction population change in a nonequilibrium steady state as $\Delta n_{c,\vecb{k}}= \left [n_{c, \vecb{k}'+e\vecb{A}(t)/\hbar}(t)- f^{FD}(\epsilon_{c,\vecb{k}'+e\vecb A (t)/\hbar})\right ]_{\vecb k'+e\vecb A(t)/\hbar =\vecb k}$. Figure~\ref{fig:pop_f}~(b-d) shows the field-induced conduction population $\Delta n_{c,\vecb{k}}$ for different field strengths, (b)~$0.01$~MV/m, (c)~$3$~MV/m, and (d)~$10$~MV/m. As seen from Fig.~\ref{fig:pop_f}~(b), the field-induced population modification is induced around the ring-shaped line, which is defined with the single-particle energy $\epsilon_{b\vecb k}$ and the Fermi energy $\epsilon_{\mathrm F}=\mu\big |_{T_e=0}$ as $\epsilon_{b\vecb k}=\epsilon_{\mathrm F}$. The population modulation is induced around the Fermi energy by the weak field excitation, and the ring structure is formed due to the circular symmetry of the Dirac cone. The increase and decrease of the conduction population $\Delta n_{c,\vecb{k}}$ show symmetric distribution along the direction of the field ($x$-axis) in the weak field regime. By contrast, the increase and decrease in population distribution become non-symmetric in the strong-field regime. As seen from Figs.~\ref{fig:pop_f}~(c) and (d), the population increase (red color region) is caused in a wider range on the left side of the Dirac point, while the population decrease (blue color region) is caused in a narrower region on the right side. The significant elongation of the population increase along the field direction can be understood as the field-induced intraband acceleration in the Brillouin zone, while the localized population decrease around the Dirac point can be understood as the field-induced displacement of the initially localized electrons around the Dirac point in Fig.~\ref{fig:pop_f}~(a).

In the previous work \cite{sato2021nonlinear}, the reduction of the conductivity has been understood as the saturation of the population imbalance around the Dirac point. To assess this interpretation, we show the conduction population distribution $n_{c,\vecb k'+e\vecb A(t)/\hbar}\big |_{\vecb k'+e\vecb A(t)/\hbar=\vecb k}$ in Fig.~\ref{fig:pop_f}~(e) instead of the population change $\Delta n_{c,\vecb{k}}$. Here, we set the field strength $E_0$ to $10$~MV/m. Note that the summation of the density in Fig.~\ref{fig:pop_f}~(a) and the density change in Fig.~\ref{fig:pop_f}~(d) corresponds to the density in Fig.~\ref{fig:pop_f}~(e). As seen from Fig.~\ref{fig:pop_f}~(e), most of the conduction population is transferred from the right side of the Dirac cone to the left side. This indicates that the population imbalance around the Dirac cone is already closely maximized and saturated since no more population can be transferred from the right side to the left side. Hence, the population imbalance cannot significantly increase more in the strong-field regime even if the field strength becomes stronger. The saturation of the population imbalance further causes the saturation of the intraband current, which is the dominant component of the current in the nonequilibrium steady-state, resulting in the reduction of the conductivity in the strong-field regime.

\section{Comparison with thermodynamic model \label{sec:thermo}}

Having established the microscopic understanding of the THz-induced HHG in graphene based on the nonequilibrium steady-state, we then study the role of the nonequilibrium nature of THz-induced electron dynamics in graphene by comparing it with the previously developed thermodynamics model \cite{mics2015thermodynamic}. In contrast to the present nonequilibrium model, the thermodynamic model is based on the thermal Fermi--Dirac distribution to describe laser-excited electronic systems under the assumption that electrons are rapidly thermalized and can be well treated as an equilibrium state with a high electron temperature $T_e$.

While equilibrium states of the thermodynamic model are characterized by the electron temperature $T_e$, nonequilibrium steady-states of the model developed in this work are naturally characterized by the applied field strength $E_0$ without relying on the temperature. To fairly compare the nonequilibrium model with the thermodynamic model, one needs to connect the electron temperature $T_e$ to the field strength $E_0$. For this purpose, we introduce the field-induced excess energy of each model. The total energy of the electronic system can be evaluated as
\begin{eqnarray}
E_{\mathrm{tot}}(t)=\frac{2}{(2\pi)^2} \int d\vecb k \mathrm{Tr}\left[H_{\vecb k+ e\vecb A(t)/\hbar} \rho_{\vecb k}(t)\right].
\label{eqn:totalenergy}
\end{eqnarray}
Then, we define the field-induced excess energy of the nonequilibrium steady-state as
\begin{eqnarray}
\Delta E^\mathrm{NEQ}_{\mathrm{excess}}(E_0)=\lim_{t\rightarrow \infty} \left [E_{\mathrm{tot}}(t)
-E_{\mathrm{tot}}(-t) \right ],
\label{eq:excess-energy-neq}
\end{eqnarray}
where $\lim_{t\rightarrow \infty} E_{\mathrm{tot}}(t)$ corresponds to the total energy in the nonequilibrium steady-state under the presence of the field, $E_0$, while $\lim_{t\rightarrow \infty} E_{\mathrm{tot}}(-t)$ corresponds to that of the equilibrium state without the field. Hence, the field-induced excess energy of the nonequilibrium model is defined as the energy difference between the nonequilibrium steady-state under an external field $E_0$ and the field-free equilibrium state.

We define the field-induced excess energy of the thermodynamic model as the energy difference between finite temperature states at $T_e$ and $300$~K, which is the initial temperature of the present nonequilibrium model:
\begin{align}
\Delta E^\mathrm{TM}_{\mathrm{excess}}&=\sum_{b=v,c}\frac{2}{(2\pi)^2}\int d\vecb k
\epsilon_{b\vecb k} \nonumber \\
& \times \left [
f^{\mathrm{FD}}\left (\epsilon_{b\vecb k},T_e,\mu \right )
-f^{\mathrm{FD}}\left (\epsilon_{b\vecb k},T_e=300~\mathrm{K},\mu \right )
 \right ].
\label{eq:excess-energy-tm}
\end{align}
Hence, $\Delta E^\mathrm{TM}_{\mathrm{excess}}$ is a function of the electron temperature $T_e$.

\begin{figure}[htb]
\includegraphics[width=1.0\linewidth]{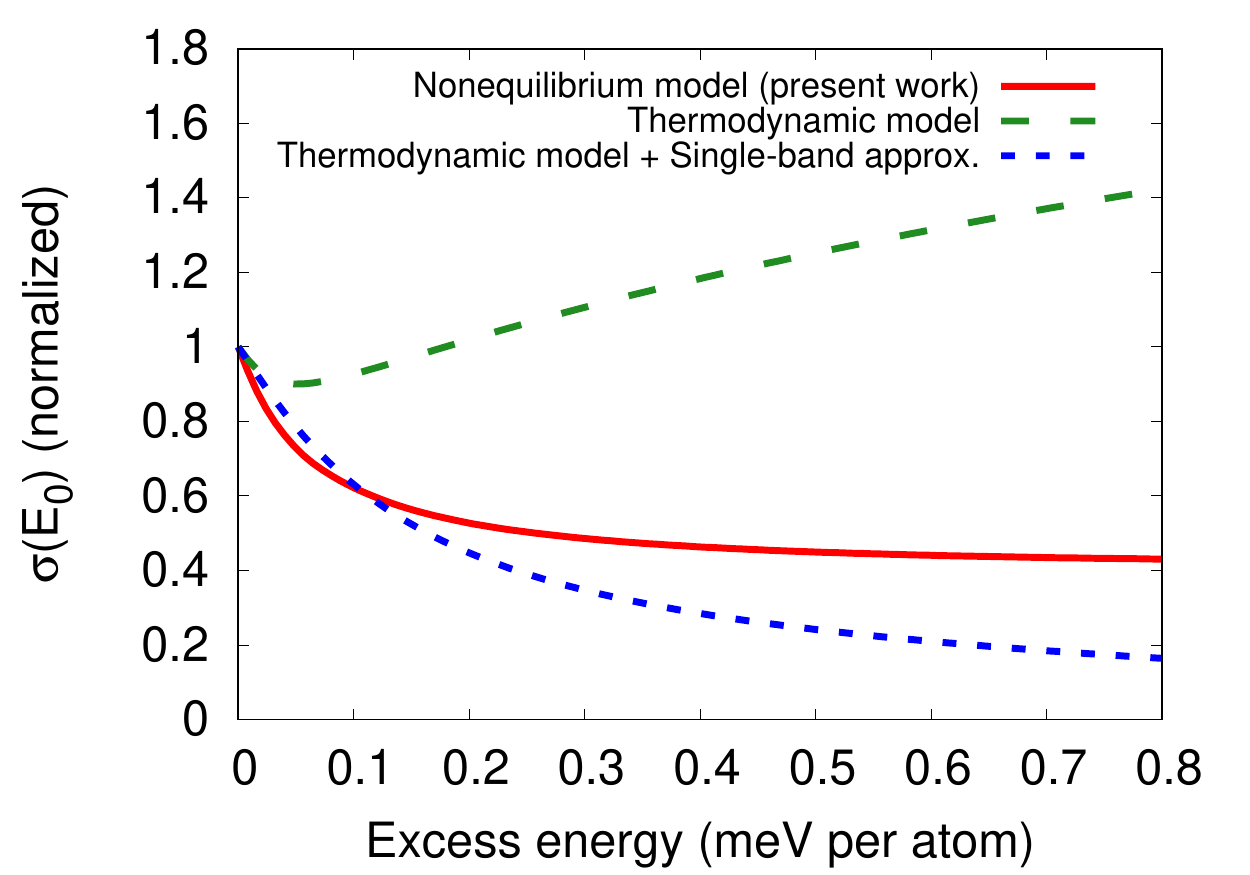}
\caption{\label{fig:compare} Computed effective conductivities are shown as a function of the excess energy. The results for the nonequilibrium steady-state (red-solid), the thermodynamic model (green-dashed), and the thermodynamic model plus the single-band approximation (blue-dotted) are shown.}
\end{figure}

With Eq.~(\ref{eq:excess-energy-neq}) and Eq.~(\ref{eq:excess-energy-tm}), the applied field strength $E_0$ to the nonequilibrium steady-state and the electron temperature $T_e$ of the thermodynamic model are connected via the excess energy. On the basis of this connection, we compare the effective conductivity $\sigma(E_0)$ of the nonequilibrium steady-state and the linear conductivity of the thermodynamic model. Figure~\ref{fig:compare} shows the conductivities of the nonequilibrium steady-state (red-solid line) and the thermodynamic model (green-dashed line). The results of the nonequilibrium steady-state are computed by setting the chemical potential $\mu$ to $170$~meV and the electron temperature $T_e$ in the relaxation operator to $300$~K. The linear conductivity of the thermodynamic model is evaluated by applying a weak field so that the induced current is described as a linear response. The results of the thermodynamic model are computed by changing the electron temperature $T_e$ but fixing the total population,
\begin{align}
N_{\mathrm{tot}}=\frac{2}{(2\pi)^2}\sum_{b=v,c} \int d\vecb k f^{FD}(\epsilon_{b\vecb k},T_e,\mu),
\label{eq:tot-pop}
\end{align}
to the value at $T_e=300$~K and $\mu=170$~meV. Hence, the chemical potential changes with the electron temperature.

As seen from Fig.~\ref{fig:compare}, the conductivity of the thermodynamic model (green-dashed line) first decreases with the increase in the excess energy, and then it increases significantly once the excess energy reaches a moderately large value. By contrast, the conductivity of the nonequilibrium steady-state (red-solid line) decreases with the increase in the excess energy in the whole investigated range. Note that the conductivity of the nonequilibrium steady-state in Fig.~\ref{fig:compare} is identical to that in Fig.~\ref{fig:conductivity} with the converted $x$-axis. The qualitative difference between the conductivities of the nonequilibrium steady-state and the thermodynamic model originates from the temperature-induced interband excitation. In the thermodynamic model, electrons are thermally excited from the valence band to the conduction band, and the number of effective carriers increases with the increase in the electron temperature, resulting in the enhancement of the conductivity. On the other hand, in the nonequilibrium steady-state, the field-induced interband excitation from the valence band to the conduction band is significantly suppressed by the Pauli blocking due to the presence of electrons in the conduction band, preventing the spurious increase in the effective carrier population and the enhancement of the conductivity.

In the previous work \cite{mics2015thermodynamic}, the microscopic mechanism of the THz-induced high-order harmonic generation and the field-induced transparency of graphene has been investigated with the thermodynamic model. The authors reported that the thermodynamic model with a single-band approximation, where only the conduction band is considered while the valence band is frozen, well reproduces the experimental results. On the other hand, it has been reported that the thermodynamic model with the two-band approximation, where both valence and conduction bands are considered, fails to reproduce the experimental results \cite{kovalev2021electrical}. Although one may naturally expect that the two-band approximation should be more accurate than the single-band approximation, the single-band approximation apparently provides a better description in the thermodynamic model. To understand the role of the single-band approximation in the thermodynamic model, we extend the above comparison between the thermodynamic model and the nonequilibrium steady-state by introducing the single-band approximation into our analysis. For this purpose, we phenomenologically freeze the population in the valence band while we continue employing the Fermi--Dirac distribution for the conduction band by modifying the Fermi--Dirac distribution as
\begin{align}
\tilde f^{\mathrm{MFD}}(\epsilon, T_e, \mu)=f^{\mathrm{FD}}(\epsilon, T_e, \mu)\Theta(\epsilon)+\Theta(-\epsilon),
\label{eq:mod-fd-dist}
\end{align}
where $\Theta(\epsilon)$ is the Heaviside step function. By replacing the Fermi--Dirac distribution of Eq.~(\ref{eq:fd-dist}) with the modified distribution of Eq.~(\ref{eq:mod-fd-dist}), we repeat the conductivity analysis with the thermodynamic model. The computed results of the thermodynamic model with the single-band approximation are shown as the blue-dotted line in Fig.~\ref{fig:compare}. One sees that the conductivity of the thermodynamic model with the single-band approximation fairly reproduces that of the nonequilibrium steady-state, showing the monotonic decrease with the increase in the excess energy. By comparing the single-band approximation with the two-band approximation in the thermodynamic model, the increase of the conductivity in the two-band approximation is significantly suppressed in the single-band approximation. This indicates that the phenomenological freezing of the valence band in the single-band approximation suppresses the spurious interband excitation in the thermodynamic model, resulting in a better description of the conductivity. By contrast, the nonequilibrium steady-state with the fully dynamical model naturally describes the suppression of the interband excitation, providing the correct behaviors of the conductivity. Therefore, the nonequilibrium steady-state picture can provide the correct description of electron dynamics in graphene under THz fields by properly taking into account both valence and conduction bands without phenomenological freezing of the valence band.

\section{Summary \label{sec:summary}} 

We developed the theoretical modeling of THz-induced electron dynamics in graphene and the high-order harmonic generation based on the quantum master equation with the relaxation time approximation. As a result of electron dynamics calculation under THz-fields, we found that the emitted harmonics are enhanced by increasing chemical potential. This theoretical finding is consistent with the recent experimental observation, where the high-order harmonic generation is enhanced by applying the gate bias voltage \cite{kovalev2021electrical}.

To develop the microscopic insight into the THz-induced electron dynamics in graphene, we introduced the nonequilibrium steady-state picture based on the quasi-static approximation. We confirmed that the THz-induced high-order harmonic generation in graphene can be well described by the quasi-static approximation, demonstrating that the nonequilibrium steady-state reflects the important aspect of the THz-induced electron dynamics. The microscopic analysis has been performed to study the role of the intraband current and the population distribution in the Brillouin zone in the steady-state. We found that the effective conductivity of graphene is significantly reduced in the strong-field regime due to the saturation of the population imbalance in the Brillouin zone. The reduction of the conductivity is consistent with the experimentally observed THz-induced transparency of graphene \cite{Hwang2013,Paul_2013,doi:10.1063/1.4902999} and the previous theoretical investigation \cite{sato2021nonlinear}. Furthermore, we found that the reduction of the effective conductivity causes the nonlinear current in the strong-field regime, resulting in the high-order harmonic generation in graphene. Therefore, the origin of the high-order harmonic generation can be understood as the saturation of the population displacement in the Brillouin zone in the strong-field regime from the viewpoint of the nonequilibrium electron dynamics.

In a recent study, the THz-induced electron dynamics in graphene have been modeled with the thermodynamic picture \cite{mics2015thermodynamic}. In contrast, we modeled the electron dynamics with the nonequilibrium picture in this work. To elucidate the role of the nonequilibrium nature of the dynamics, we further investigated both the thermodynamic picture and the nonequilibrium picture. As a result, we found that the thermodynamic model shows a spurious enhancement of the electric conductivity under the irradiation of strong THz fields due to the significant interband transitions from the valence to conduction bands. By employing the single-band approximation introduced in the previous work \cite{kovalev2021electrical}, we artificially froze the valence band and further computed the conductivity with the thermodynamic model. Consistently with the previous work \cite{kovalev2021electrical}, the single-band approximation suppresses the spurious interband excitation, and the computed conductivity in the thermodynamic picture properly shows the decreasing trend under the field irradiation, which is consistent with the experimental observation of the field-induced transparency of graphene \cite{Hwang2013,Paul_2013,doi:10.1063/1.4902999}. By contrast to the thermodynamic model, the nonequilibrium model developed in this work properly describes the decreasing trend of the conductivity under the field irradiation even without artificially freezing the valence band. This indicates that the nonequilibrium nature of electron dynamics is essential to describe the reduction of the conductivity under the field irradiation by preventing the spurious interband excitation from the viewpoint of the comparison with the thermodynamic model. The fully dynamical calculation based on the quantum master equation offers a natural description of the nonequilibrium nature of field-induced phenomena such as symmetry breaking and delayed responses. The theoretical studies on these aspects of the nonequilibrium nature of light-induced phenomena are already underway.

\begin{acknowledgments}
This work was supported by JSPS KAKENHI Grant Numbers JP20K14382 and JP21H01842, the Cluster of Excellence 'Advanced Imaging of Matter' (AIM), Grupos Consolidados (IT1249-19) and Deutsche Forschungsgemeinschaft (DFG) --SFB-925-- project 170620586. The Flatiron Institute is a division of the Simons Foundation.
\end{acknowledgments}

\bibliography{ref}

\appendix
\section{High-order harmonic generation in the quasi-static picture \label{appendix:qs-picture}}

Here, we describe a method to compute the high-order harmonic generation in the quasi-static approximation. The spectra of the high-order harmonic generation are computed with the Fourier transform of the induced current with Eq.~(\ref{eqn:spectrum}). In the quasi-static approximation, we approximate the induced current $\vecb J(t)$ with the steady current $\vecb J_S(E_0)$ in Eq.~(\ref{eq:steady-current}) by substituting the instantaneous electric field as
\begin{align}
\vecb J(t)\approx \vecb J_S\left( \vecb E(t) \right).
\label{eq:appendix-steady-current}
\end{align}

\begin{figure}[htb]
\includegraphics[width=1\linewidth]{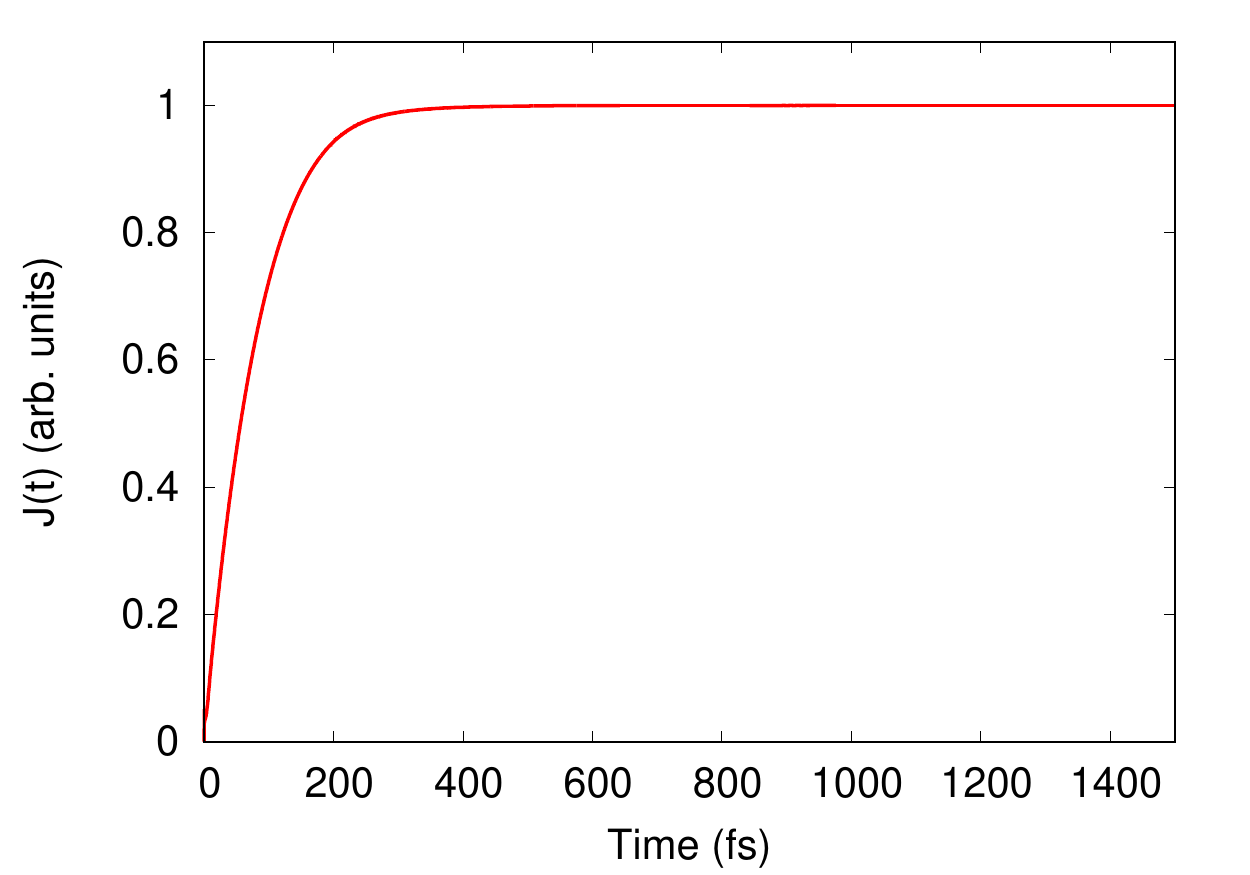}
\caption{\label{fig:steady}
Electric current in graphene under a static electric field, $E_0=8.5$~MV/m.}
\end{figure}

To evaluate the approximated current in Eq.~(\ref{eq:appendix-steady-current}), we first evaluate the steady current in Eq.~(\ref{eq:steady-current}) for several field strengths. For practical evaluation, we compute the electron dynamics under a static electric field, $\vecb E_0=E_0\vecb e_x$. Figure~\ref{fig:steady} shows the evaluated current under a static field as a function of time. In this simulation, the chemical potential $\mu$ is set to $170$~meV, and the field strength $E_0$ is set to $8.5$~MV/m. The initial state at $t=0$ is set to the thermal equilibrium state. As seen from Fig.~\ref{fig:steady}, the electric current is induced at $t=0$ due to the field application, and it reaches a value of the steady-state, $\vecb J_S(E_0)$. Therefore, we confirm that the electronic system evolved with Eq.~(\ref{eqn:masterequation}) under a static electric field reaches a nonequilibrium steady-state after sufficiently long time propagation.

\begin{figure}[htb]
\includegraphics[width=1.1\linewidth]{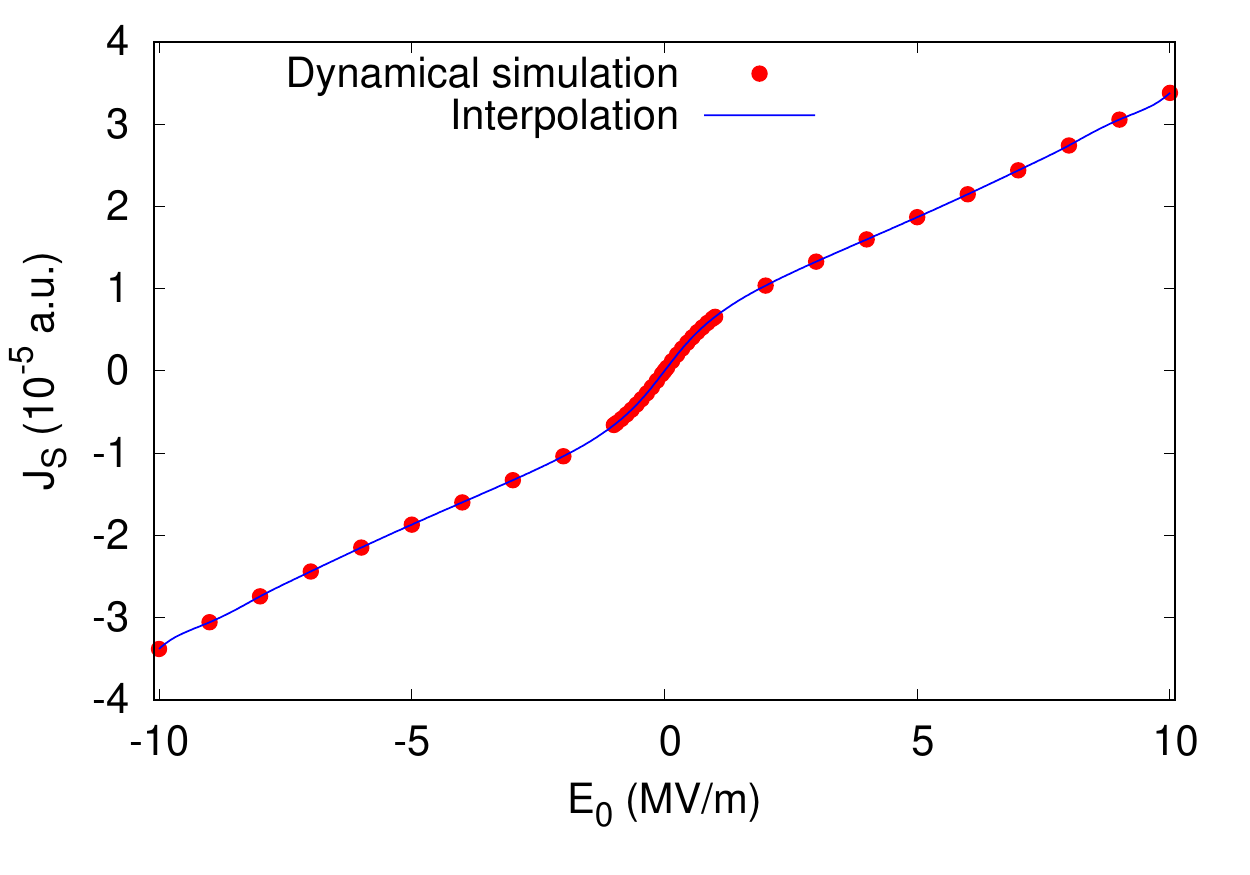}
\caption{\label{fig:insert} 
Steady current $\vecb J_S(E_0)$ as a function of field strength $E_0$. The results of the fully dynamical calculation are showns as the red points, while the interpolated result is shown as the blue-solid line.}
\end{figure}

We repeat the above simulations by changing the field strength $E_0$ and evaluating the values of the steady current. We denote the $k$th set of the employed field strength and the evaluated current as $E_k$ and $\vecb J_k$, respectively. The computed steady current $\vecb J_k$ is shown as the red points in Figure~\ref{fig:insert} as a function of the applied field strength $E_k$. To construct the continuous function of the steady current $\vecb J_S(E_0)$ from the discrete data points $\{E_k, \vecb J_k\}$ in Fig.~\ref{fig:insert}, we employ the following two-step interpolation procedure.

As the first step to construct the continuous function, we perform a polynomial regression with the following odd function
\begin{align}
\vecb{J}_{\mathrm{polynomials }}(E_0)= \sum\limits_{j=0}^{4} \vecb e_x \alpha^{(2j+1)} E^{2j+1}_0,
\label{eq:appendix-polynomial}
\end{align}
where $\alpha^{(j)}$ are optimization parameters. These parameters are optimized so that the polynomial function $\vecb{J}_{\mathrm{polynomials }}(E_0)$ well reproduces the discrete points $\{E_k, \vecb J_k\}$ in Fig.~\ref{fig:insert}.

As the second step, we refine the discrepancy between the discrete points in Fig.~\ref{fig:insert} and the polynomial function $\vecb{J}_{\mathrm{polynomials }}(E_0)$. Practically, we first define the residual error of the above polynomial regression as
\begin{align}
\Delta \vecb{J}_k=\vecb{J}_{k}-\vecb{J}_{\mathrm{polynomials}}(E_k).
\end{align}
Then, we apply the spline interpolation to the data points $\{E_k, \Delta \vecb{J}_k\}$. Here, we denote the interpolated function as $\Delta \vecb{J}_{\mathrm{spline}}(E_0)$. Finally, we approximate the continuous function, $\vecb J_S(E_0)$, as
\begin{align}
\vecb J_S(E_0)\approx \vecb{J}_{\mathrm{polynomials }}(E_0)+\Delta \vecb{J}_{\mathrm{spline}}(E_0).
\label{eqn:approx}
\end{align}

\begin{figure}[htb]
\includegraphics[width=1.0\linewidth]{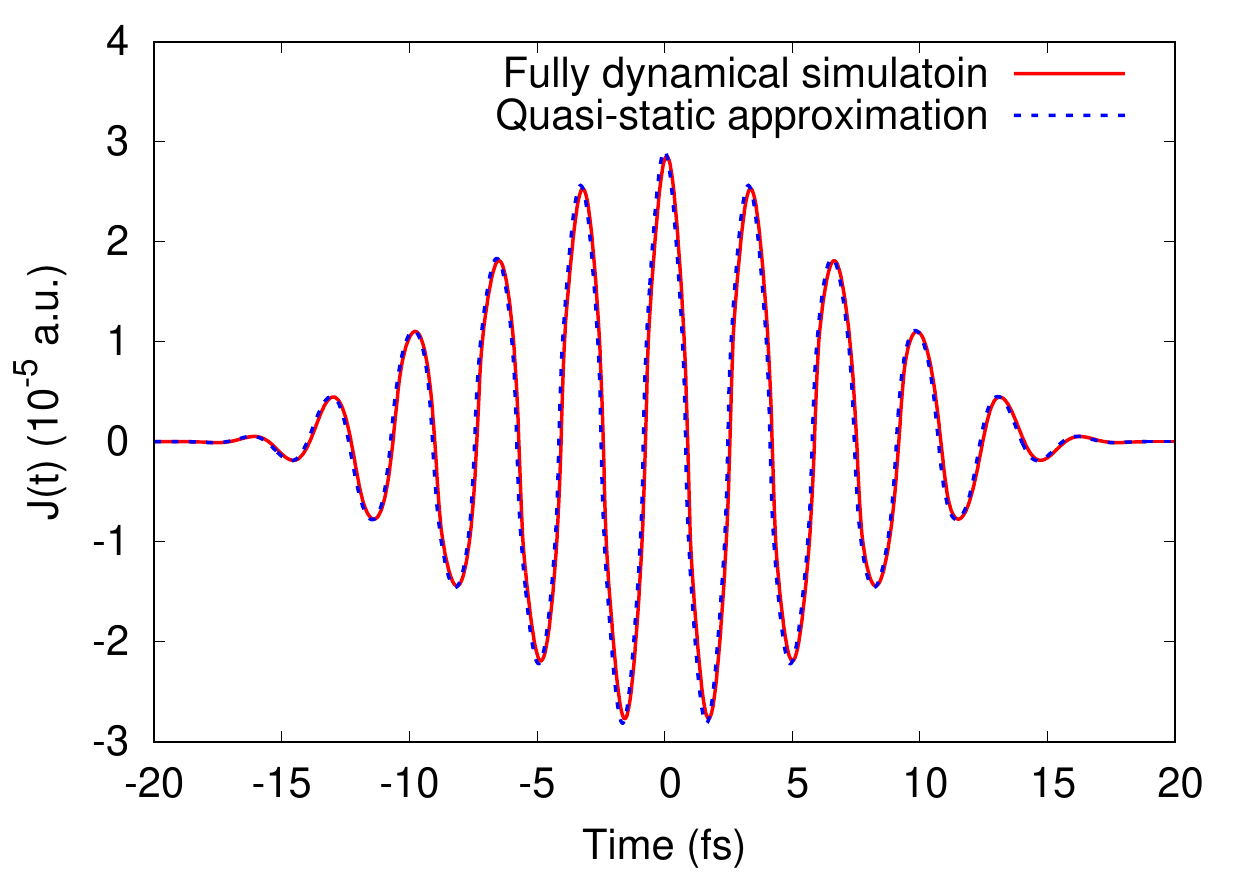}
\caption{\label{fig:current} Comparison of the THz-induced current computed with the fully dynamical calculation and the quasi-static approximation.}
\end{figure}

By employing the approximated function, Eq.~(\ref{eqn:approx}), we evaluate the THz-induced electric current with the quasi-static approximation, Eq.~(\ref{eq:appendix-steady-current}). Figure~\ref{fig:current} shows the computed current as a function of time with the quasi-static approximation. For comparison, the result of the fully dynamical calculation is also shown. By applying the Fourier transform to the obtained current in Fig.~\ref{fig:current}, we obtain the HHG spectra in Fig.\ref{fig:hhg_mu}~(b).

\end{document}